\documentclass[fleqn]{annalen}
\usepackage{graphics}
\begin{document}
\newcommand{\volume}{8}              
\newcommand{\xyear}{1999}            
\newcommand{\issue}{5}               
\newcommand{\recdate}{29 July 1999}  
\newcommand{\revdate}{dd.mm.yyyy}    
\newcommand{\revnum}{0}              
\newcommand{\accdate}{dd.mm.yyyy}    
\newcommand{\coeditor}{ue}           
\newcommand{\firstpage}{507}         
\newcommand{\lastpage}{510}          
\setcounter{page}{\firstpage}        
\newcommand{\keywords}{spin-$\frac{1}{2}$ $XY$ chain, 
correlated disorder,
density of states} 
\newcommand{\PACS}{75.10.--b}
\newcommand{\shorttitle}{O. Derzhko et al., 
Thermodynamics of the $s=\frac{1}{2}$ transverse $XX$ chain} 
\title{Thermodynamics of the $s=\frac{1}{2}$ transverse $XX$ chain\\ 
with Dzyaloshinskii--Moriya interaction\\
in the presence of correlated Lorentzian disorder}
\author{O.\ Derzhko$^{1,2}$, J.\ Richter$^{3}$, 
and V.\ Derzhko$^{2}$} 
\newcommand{\address}
  {$^{1}$Institute for Condensed Matter Physics, 
  1 Svientsitskii St., L'viv--11, 290011, Ukraine\\ 
  $^{2}$Chair of Theoretical Physics,
  Ivan Franko State University of L'viv,\\
  \hspace*{0.5mm} 
  12 Drahomanov St., 
  L'viv--5, 290005, Ukraine\\
  $^{3}$Institut f\"{u}r Theoretische Physik,
  Universit\"{a}t Magdeburg,\\
  \hspace*{0.5mm} 
  P.O. Box 4120, D--39016 Magdeburg, Germany}
\newcommand{\email}{\tt derzhko@icmp.lviv.ua} 
\maketitle
\begin{abstract}
The spin-$\frac{1}{2}$ transverse $XY$ chain with correlated disorder may
exhibit the nonzero averaged transverse magnetization at zero averaged 
transverse field. We discuss this peculiarity connected with asymmetry in 
the density of states and induced by the correlated disorder analysing 
the moments of the density of states.
\end{abstract}

\vspace{10mm}

Recently the thermodynamic properties of the 
spin-$\frac{1}{2}$ transverse $XY$ chain with correlated disorder
have been discussed \cite{001,002}.
The model consists of
$N\to\infty$ spins $\frac{1}{2}$
on a circle governed by the Hamiltonian
\begin{eqnarray}
\label{001}
H=\sum_{n=1}^{N}\Omega_ns_n^z
+\sum_{n=1}^{N}J_n\left(s^x_ns^x_{n+1}+s^y_ns^y_{n+1}\right)
+D\sum_{n=1}^{N}\left(s^x_ns^y_{n+1}-s^y_ns^x_{n+1}\right)
\nonumber\\
=\sum_{n=1}^{N}\Omega_n\left(s^+_ns^-_n-\frac{1}{2}\right)
+\sum_{n=1}^{N}
\left(
\left(
{\cal{J}}_n+i{\cal{D}}
\right)
s^+_ns^-_{n+1}
+\left(
{\cal{J}}_n-i{\cal{D}}
\right)
s^-_ns^+_{n+1}
\right).
\end{eqnarray}
The isotropic interactions between neighbouring sites 
$J_n=2{\cal{J}}_n$
are assumed to
be independent random variables
each with probability distribution
$p(J_n)$
and the transverse field at site $\Omega_n$
is determined  by the surrounding couplings $J_{n-1}$ and $J_n$
according to the formula
\begin{eqnarray}
\label{002}
\Omega_n=\overline{\Omega}
+\frac{a}{2}\left(J_{n-1}+J_n-2\overline{J}\right).
\end{eqnarray}
Here $\overline{\Omega}$ and $\overline{J}$ are the mean values of 
$\Omega_n$ and $J_n$ respectively
and $a$ is a real parameter.
Besides the isotropic interaction the model includes the 
Dzyaloshinskii--Moriya interaction between neighbouring sites 
$D=2{\cal{D}}$.
Due to the 
imposed relation between the transverse field and random
exchange couplings (\ref{002})
that is a model of correlated off--diagonal and 
diagonal disorder.
For the case of the Lorentzian probability distribution
$p(J_n)=\frac{1}{\pi}
\frac{\Gamma}{\left(J_n-\overline{J}\right)^2+\Gamma^2}$
and $\vert a\vert\ge 1$
the Jordan--Wigner method \cite{003}
and the method elaborated by John and Schreiber \cite{004} 
permits one to derive an explicit expression for the density of magnon
states 
$\overline{\rho(E)}$ 
($\overline{(\ldots)}
\equiv\ldots\int_{-\infty}^{\infty}dJ_np(J_n)\ldots (\ldots)$)
and hence to study rigorously the thermodynamic 
properties of the model (\ref{001}), (\ref{002})
\cite{001}.
Apparently the most interesting property of that model 
caused by the correlated disorder is the asymmetry in the density of states
$\overline{\rho(E-\Omega_0)}\ne\overline{\rho(-E+\Omega_0)}$ 
especially nicely pronounced for $\vert a\vert\approx 1$
that has a number of intriguing consequences for thermodynamics. For 
example, the considered model may exhibit a nonzero averaged transverse 
magnetization 
$\overline{m_z}=-\frac{1}{2}\int_{-\infty}^{\infty}dE
\overline{\rho(E)}\tanh\frac{E}{2kT}\ne 0$
at zero averaged transverse field $\overline{\Omega}=0$
and $T=0$ 
since in such a limit 
$\overline{m_z}
=\frac{1}{2}\int_{-\infty}^{0}dE\overline{\rho(E)}
-\frac{1}{2}\int_{0}^{\infty}dE\overline{\rho(E)}$.
These findings were also confirmed numerically
for other types of correlated (not necessarily Lorentzian) disorder 
\cite{002}.
It was also shown that the Dzyaloshinskii--Moriya interaction may lead to a 
decrease of the nonzero $\overline{m_z}$ 
at zero $\overline{\Omega}$ that appears 
because of the correlated disorder \cite{001}.
In the present paper these somewhat unusual properties 
are discussed
in more detail. In particular, we calculate 
rigorously the moments of 
the density of states 
$\overline{M^{(p)}}\equiv\int_{-\infty}^{\infty}dE
E^{p}\overline{\rho(E)}$
revealing the influence of the correlated disorder on the asymmetry in the 
density of states. 

Let us consider at first more closely the mentioned magnetic property,  
i.e., because of randomness the model may exhibit a nonzero averaged 
magnetization at the zero averaged field. This peculiarity is conditioned by 
the introduced {\em correlated} disorder. Consider a certain random 
realization of the chain defined by 
(\ref{001}), (\ref{002}), for example, with Lorentzian probability 
distribution.
One may expect that there will be the same numbers of sites surrounded by 
stronger than $\overline{J}$ isotropic couplings
as the sites surrounded by weaker than 
$\overline{J}$ isotropic couplings.
Because of the adopted relation (\ref{002}) 
for $\overline{\Omega}=0$
the transverse fields at the 
former and the latter sites 
have the same values but the opposite directions so that 
$\sum_{n=1}^N\Omega_n=0$. Bearing in mind a corresponding 
classical chain (\ref{001}) 
in which the intersite interaction tries to arrange spins in $xy$ plane 
whereas the transverse field tries to align spins along $z$ axis
one may expect that the sites surrounded by strong 
couplings exhibit smaller magnetization whereas the sites surrounded by the 
weak couplings exhibit larger magnetization in the opposite direction. As a 
result 
$\frac{1}{N}\sum_{n=1}^N\langle s_n^z\rangle\ne 0$.
With increase of $\vert a\vert$ the difference in that magnetizations 
becomes smaller. 
Switching on an additional constant coupling $D$ 
also 
makes that difference 
smaller. Thus, a nonzero $\overline{m_z}$ at zero 
$\overline{\Omega}$ appears because of the introduced relation between 
the isotropic couplings and the on--site transverse fields.

Let us next examine the asymmetry in the density of states more 
quantitatively calculating for this purpose the moments of the density of 
states. They can be written as follows \cite{005}
\begin{eqnarray}
\label{003}
M^{(p)}=\frac{1}{N}\sum_{n=1}^N
\left\langle
\left\{
\left[\ldots\left[
c_n,H
\right],\ldots, H\right], c_n^+
\right\}
\right\rangle
\end{eqnarray}
where
\begin{eqnarray}
\label{004}
H=\sum_{n=1}^{N}\Omega_n
\left(c_n^+c_n-\frac{1}{2}\right)
+\sum_{n=1}^{N}
\left(
\left({\cal{J}}_n+i{\cal{D}}\right)c^+_nc_{n+1}
-\left({\cal{J}}_n-i{\cal{D}}\right)c_nc^+_{n+1}
\right)
\end{eqnarray}
is the Hamiltonian of the nonuniform spin chain 
(\ref{001})
in the Jordan--Wigner 
picture. 
Evidently, $M^{(p)}$ 
can be calculated exactly
for an arbitrary nonuniform spin-$\frac{1}{2}$ 
transverse $XY$ chain.
From (\ref{003}), (\ref{004}) one obtains  
\begin{eqnarray}
\label{005}
M^{(1)}=\frac{1}{N}\sum_{n=1}^N\Omega_n,
\;\;\;\;\;
M^{(2)}=\frac{1}{N}\sum_{n=1}^N
\left(
\Omega_n^2+{\cal{J}}_{n-1}^2+{\cal{J}}_{n}^2+2{\cal{D}}^2
\right),
\nonumber\\
M^{(3)}=\frac{1}{N}\sum_{n=1}^N
\left(
\Omega_{n}^3
+\left(\Omega_{n-1}+2\Omega_n\right)
\left({\cal{J}}_{n-1}^2+{\cal{D}}^2\right)
\right.
\nonumber\\
\left.
+\left(2\Omega_{n}+\Omega_{n+1}\right)
\left({\cal{J}}_{n}^2+{\cal{D}}^2\right)
\right).
\end{eqnarray}

Consider now random spin-$\frac{1}{2}$ transverse $XY$ chains.
For correlated disorder (\ref{002}) Eq. (\ref{005}) yields
\begin{eqnarray}
\label{006}
{\overline{M^{(1)}}}
=\overline{\Omega},
\;\;\;\;\;
{\overline{M^{(2)}}}
=\overline{\Omega}^2
+\frac{1}{2}\left(\overline{J^2}+D^2\right)
+\frac{a^2}{2}\left(\overline{J^2}-\overline{J}^2\right),
\nonumber\\
{\overline{M^{(3)}}}
=\overline{\Omega}^3
+\frac{3}{2}\overline{\Omega}\left(\overline{J^2}+D^2\right)
+\frac{3a^2}{2}
\overline{\Omega}\left(\overline{J^2}-\overline{J}^2\right)
\nonumber\\
+\frac{a^3}{4}\left(\overline{J^3}
-3\overline{J}\;{\overline{J^2}}+2\overline{J}^3\right)
+\frac{3a}{4}\left(\overline{J^3}-\overline{J}\;{\overline{J^2}}\right),
\end{eqnarray}
whereas
for not correlated off--diagonal and diagonal disorder 
($\Omega_n$ are independent random variables) 
instead of (\ref{006}) one gets
\begin{eqnarray}
\label{007}
{\overline{M^{(1)}}}
=\overline{\Omega},
\;\;\;\;\;
{\overline{M^{(2)}}}
={\overline{\Omega^2}}
+\frac{1}{2}\left({\overline{J^2}}+D^2\right),
\;\;\;\;\;
{\overline{M^{(3)}}}
={\overline{\Omega^3}}
+\frac{3}{2}\overline{\Omega}\left(\overline{J^2}+D^2\right).
\end{eqnarray}
For $\overline{\Omega}=0$ from (\ref{007}) follows that 
${\overline{M^{(3)}}}=0$
(e.g., 
for the rectangle probability distribution for $J_n$ and $\Omega_n$,
i.e. 
$p(x)=\frac{1}{2\Delta}$ 
if 
$\overline{x}-\Delta\le x\le\overline{x}+\Delta$
and 
$p(x)=0$  otherwise,
${\overline{M^{(1)}}}=\overline{\Omega}$,
${\overline{M^{(2)}}}=\overline{\Omega}^2
+\frac{1}{3}\Delta^2
+\frac{1}{2}(\overline{J}^2+D^2+\frac{1}{3}\Delta^2)$,
${\overline{M^{(3)}}}=\overline{\Omega}^3
+\overline{\Omega}\Delta^2
+\frac{3}{2}\overline{\Omega}
(\overline{J}^2+D^2+\frac{1}{3}\Delta^2)$)
as it should be if
$\overline{\rho(E)}=\overline{\rho(-E)}$.
Contrary,
for $\overline{\Omega}=0$
from (\ref{006}) follows that
${\overline{M^{(3)}}}\ne 0$
(e.g.,
for the rectangle probability distribution 
${\overline{M^{(1)}}}=\overline{\Omega}$,
${\overline{M^{(2)}}}=\overline{\Omega}^2
+\frac{1}{2}(\overline{J}^2+D^2+\frac{1}{3}\Delta^2)
+\frac{a^2}{6}\Delta^2$,
${\overline{M^{(3)}}}=\overline{\Omega}^3
+\frac{3}{2}\overline{\Omega}
(\overline{J}^2+D^2+\frac{1}{3}\Delta^2)
+\frac{a^2}{2}\overline{\Omega}\Delta^2
+\frac{a}{2}\overline{J}\Delta^2$)
that demonstrate explicitly the appearance of the asymmetry 
in $\overline{\rho(E)}$ due to the 
correlated disorder.
One can also see from (\ref{006}) that for $\overline{\Omega}=0$ 
${\overline{M^{(3)}}}$ is an odd function of $a$, 
it vanishes as $\vert a\vert\rightarrow 0$
or in the non--random limit 
$p(J_n)=\delta(J_n-\overline{J})$ 
that is in agreement with the results obtained for the Lorentzian 
probability distribution $p(J_n)$ \cite{001}. 
Besides, for $\overline{\Omega}=0$ 
if $D$ increases 
${\overline{M^{(2)}}}$ is increased but ${\overline{M^{(3)}}}$ is not 
and hence the asymmetry in $\overline{\rho(E)}$ decreases as it was found in 
\cite{001}. Finally, from the explicit expressions for 
${\overline{M^{(p)}}}$
in the case of rectangle probability distribution 
for $\overline{\Omega}=0$ one finds that
${\overline{M^{(2)}}}\sim a^2$
and
${\overline{M^{(3)}}}\sim a$ 
that hints at the recovering of symmetry in 
$\overline{\rho(E)}$ 
while $\vert a\vert\rightarrow \infty$.

It is worthwhile to note 
that the asymmetry in the density of states may appear in 
non--random model as well. Consider, for example,
a uniform one--dimensional model of 
spinless fermions with next nearest hopping with the Hamiltonian
\begin{eqnarray}
\label{008}
H=\Omega\sum_{n=1}^{N}
\left(c_n^+c_n-\frac{1}{2}\right)
+{\cal{J}}\sum_{n=1}^{N}
\left(c^+_nc_{n+1}-c_nc^+_{n+1}\right)
\nonumber\\
+{\cal{K}}\sum_{n=1}^{N}
\left(c^+_nc_{n+2}-c_nc^+_{n+2}\right)
\end{eqnarray}
for which
one can easily obtain the following moments of the density of states
\begin{eqnarray}
\label{009}
M^{(1)}=\Omega,
\;
M^{(2)}=\Omega^2+2{\cal{J}}^2+2{\cal{K}}^2,
\;
M^{(3)}=\Omega
\left(\Omega^2+6{\cal{J}}^2+6{\cal{K}}^2\right)
+6{\cal{J}}^2{\cal{K}}.
\end{eqnarray}
According to (\ref{009})
$M^{(3)}\ne 0$ at $\Omega=0$.
One can easily construct a spin-$\frac{1}{2}$ chain which in the 
Jordan--Wigner picture corresponds to (\ref{009}), namely,
\begin{eqnarray}
\label{010}
H=\Omega\sum_{n=1}^{N}s_n^z
+2{\cal{J}}\sum_{n=1}^{N}\left(s^x_ns^x_{n+1}+s^y_ns^y_{n+1}\right)
\nonumber\\
-4{\cal{K}}\sum_{n=1}^{N}\left(s^x_ns_{n+1}^zs^x_{n+2}
+s^y_ns^z_{n+1}s^y_{n+2}\right).
\end{eqnarray}
Spin-$\frac{1}{2}$ chain (\ref{010}) similarly to the model with randomness  
defined by (\ref{001}), (\ref{002})
exhibit a nonzero transverse magnetization 
$m_z$ at the zero transverse field $\Omega$.

To summarize, we discussed the appearance of nonzero averaged magnetization 
at the zero averaged field 
in spin-$\frac{1}{2}$ transverse $XY$ chain
induced by correlated off--diagonal and diagonal 
disorder. This happens because of the asymmetry in the density of states 
$\overline{\rho(E)}$
that is indicated by the nonzero third moment of the density of states 
$\overline{M^{(3)}}$. 
We calculated explicitly $M^{(3)}$ and showed that at zero 
averaged field we have
$\overline{M^{(3)}}\ne 0$ for correlated disorder and
$\overline{M^{(3)}}= 0$ for not correlated disorder. 
The obtained few first moments of the density of states permit to examine 
rigorously the high--temperature properties of an arbitrary nonuniform 
(random) spin-$\frac{1}{2}$ $XY$ chain. 
We also constructed a uniform spin-$\frac{1}{2}$ chain with three site 
interaction which 
exhibits nonzero magnetization at zero field.

\vspace*{0.25cm} \baselineskip=10pt{\small \noindent 
This work was partly supported
by the DFG (projects 436 UKR 17/20/98 and Ri 615/6-1).
It was presented at 
the International Conference
LOCALIZATION 1999
(Hamburg, 1999).
O.~D. is grateful to the
organizers 
for the financial support.}
%
%
%
%
%
%
%
%
%
%
%
%

\end{document}